  \documentclass[useAMS,usenatbib]{mn2e}

\usepackage{graphicx}
\usepackage{subfigure}
\usepackage{color}
\usepackage[authoryear]{natbib}
\usepackage{names}

\def\eg{{\it e.g.} }

\def\ie{{\it i.e.} }

\begin{document}


\title[]{The scattering of small bodies in planetary systems: constraints on the possible orbits of cometary material.}


\author[A. Bonsor et al.]
  {A. Bonsor$^1$\thanks{Email: abonsor@ast.cam.ac.uk},
    M. C. Wyatt$^1$\\
  $^1$ Institute of Astronomy, University of Cambridge, Madingley Road,
  Cambridge CB3 0HA, UK}

\maketitle

\begin{abstract}

The scattering of small bodies by planets is an important dynamical process in planetary systems. In this paper we present an analytical model to describe this process using the simplifying assumption that each particle's dynamics is dominated by a single planet at a time. As such the scattering process can be considered as a series of three body problems during each of which the Tisserand parameter with respect to the relevant planet is conserved. This constrains the orbital parameter space into which a particle can be scattered. Such arguments have previously been applied to the process by which comets are scattered to the inner Solar System from the Kuiper belt. Our analysis generalises this for an arbitrary planetary system. For particles scattered from an outer belt directly along a chain of planets, based on the initial value of the Tisserand parameter, we find that it is possible to (i) determine which planets can eject the particles from the system, (ii) define a minimum stellar distance to which particles can be scattered, and (iii) constrain range of particle inclinations (and hence the disc height) at different distances. Applying this to the Solar System, we determine that the planets are close to optimally separated for scattering particles between them. Concerning warm dust found around stars that also have Kuiper belt analogues, we show that, if there is to be a dynamical link between the outer and inner regions, then certain architectures for the intervening planetary system are incapable of producing the observations. We speculate that the diversity in observed levels of warm dust may reflect the diversity of planetary system architectures. Furthermore we show that for certain planetary systems, comets can be scattered from an outer belt, or with fewer constraints, from an Oort cloud analogue, onto star-grazing orbits, in support of a planetary origin to the metal pollution and dustiness of some nearby white dwarfs. In order to make more concrete conclusions regarding scattering processes in such systems, it is necessary to consider not only the orbits available to scattered particles, but the probability that such particles are scattered onto the different possible orbits.


\end{abstract}

\section{Introduction}

The scattering of small bodies is an important dynamical process in many planetary systems. One classic example is the population of small bodies close to the Sun, many of which originate further out in the Solar System, from where they were scattered inwards. Near-Earth asteroids (NEAs) originate in the asteroid belt. Many left the belt after being destabilised by resonances with Jupiter and then scattered by the terrestial planets \citep{MorbiAstIII}. Visible comets are either objects scattered inwards from the Kuiper belt or the Oort cloud \citep{LevisonDuncan97}. The scattering of small bodies has not been considered in detail for extra-solar planetary systems, mainly due to the lack of constraints on the structure of the planetary system. There is, however, evidence for small bodies in many extra-solar planetary systems. Dust belts, known as debris discs are seen around hundreds of main sequence stars \citep{wyattreview}. Observations, particularly resolved images, suggest that debris discs interact with planets \citep{fomalahautresolvedscatteredlight, Moerchen10, Greaves05}, etc. Assuming a similar nature to our Solar System, it is reasonable to expect that scattering in these systems can also result in a comet-like population. The expected level and distribution of this comet population may differ substantially from the Solar System, depending on the individual planetary system architecture. 

Evidence of such a comet-like population may exist from observations of {\it warm} dust discs around a handful of main sequence stars \citep{Wyattetacorvi, Gaidos99, Beichman05, Song05}. Comets or asteroids in the position of the observed dust belts have a short lifetime against collisions and drag forces. They cannot have existed for the entire main sequence lifetime in their observed position \citep{Wyatt07hot}. One possible explanation is that the material originated in a cold, outer belt. It could be that we are observing a comet-like population, that is continuously replenished from scattering of material from the outer belt by intervening planets \citep{Wyatt07hot}. Alternatively, it could be a transient event, resulting from the stochastic collision of two larger bodies \citep{Song05}, maybe in a similar manner to the impact that formed the Earth-Moon system. Or, material could be transported inwards from the outer belt during a LHB type event \citep{Booth09} or by drag forces \citep{Eps_hotdust}.

Another piece of evidence for the scattering of material in exo-planetary systems comes from observations of evolved stars. 25\% of DA white dwarfs show unexpected metal pollution \citep{Zuckerman03}, whilst 1-3\% of DA white dwarfs with cooling ages less than 0.5Gyr have excess emission in the infra-red consistent with a close-in dust disc \citep{farihi09}. The composition of the polluting material closely resembles planets \citep{Klein10} and there is good evidence that it is not material accreted from the inter-stellar medium \citep{Farihi10ism}. The best theories, therefore, suggest that it originates in an outer planetary system \citep{JuraWD03, Gansicke06, kilic06, vonhippel07, farihi09, Farihi10, melis10}. As the star loses mass on the giant branch, dynamical instabilities can be induced in the outer planetary system \citep{DebesSigurdsson}. These can lead to comets or asteroids being scattered by interior planets onto star-grazing orbits, where they are tidally disrupted. Material from the tidally disrupted asteroids or comets forms the observed discs and accretes onto the star. The ability of evolved planetary systems to scatter comets or asteroids onto star-grazing orbits requires further detailed investigation, although previous work has considered an Oort cloud origin of the scattered bodies \eg \citep{Alcock86, DebesSigurdsson}.

In this work the scattering of small bodies in an arbitrary planetary system is investigated. N-body simulations are typically used to model such scattering \citep{LevisonDuncan97, JupiterFoF, Holman93}. A deeper understanding of the general properties of such scattering can, however, be achieved using analytical arguments. Simulations of scattered Kuiper belt objects have found that the scattering process can be approximated as a series of three-body problems, as the scattered bodies are passed from one planet to the next \citep{LevisonDuncan97}. While such particles are under the influence of one of the planets, their dynamical evolution can be approximated by the circular restricted three-body problem in which the orbits of the particles must be such that their Tisserand parameters, $T_p$, \citep{Tisserand, Murray&Dermott} are conserved, where

\begin{equation}
T_p= \frac{a_p }{a} +2\sqrt{\frac{(1-e^2)a}{a_p}} \cos (I),
\label{eq:tiss}
\end{equation}
where a, e, I are the comet's semi-major axis, eccentricity and inclination and $a_p$ is the planet's semi-major axis. 
This conservation is so fundamental to cometary dynamics that it is used to classify cometary orbits \citep{Horner03, gladmannomenclature}.

In this work we use the conservation of the Tisserand parameter to constrain the orbits of scattered particles in a planetary system with an arbitrary configuration. In section~\ref{sec:initial} we discuss how planetesimals are scattered from an outer belt, in an otherwise stable planetary system. We then outline our constraints on the orbits of particles scattered by a single planet in section~\ref{sec:onepl}, which we extend to two planets in section~\ref{sec:twopl} and arbitrarily many planets in section~\ref{sec:multi}. In section~\ref{sec:application} we consider the application of this analysis to our Solar System, systems with {\it warm} dust discs and polluted white dwarfs.



\section{Scattering of planetesimals }
\label{sec:initial}

During the planet formation process, a planet that forms in a disc of planetesimals, will swiftly clear a zone around it, both by scattering processes and resonant interactions with the planet. Analytically the size of the planet's cleared zone can be approximated. Criterion for the overlap of mean motion resonances determine a region around the planet within which orbit's are chaotic \citep{Wisdom1980}, whilst the Jacobi constant can be used to determine the zone within which orbits can be planet crossing \citep{Gladman90}. Simulations have shown that Neptune clears such a zone in less than $10^5$ yr \citep{LevDun93, Holman93}, but more generally one might expect 1,000 conjunctions for this clearing to take effect \citep{DQT89}. Material removed from this region may be ejected, whilst some fraction remains on bound, eccentric orbits, with pericentres close to the planet's orbit, forming an analogue to Neptune's {\it scattered disc}. After many scatterings some of this material may reach far enough from the star to interact with the Galactic tide \citep{tremaine93} and eventually populate an analogue of the Oort cloud.

Planetesimals outside of this zone could in principle be long term stable. However, N-body simulations of Neptune and the Kuiper belt find that Kuiper belt objects are still scattered by Neptune at late times \citep{duncan95,Holman93, LevisonDuncan97, Emelyanenko04, Morbi97}. The Kuiper belt has a complicated structure of stable and unstable regions. The gravitational effects of Neptune and the inner planets result in the overlap of secular or mean-motion resonances producing thin chaotic regions, within the otherwise stable region \citep{Kuchner02, Lykawka05} and small unstable regions within otherwise stable mean motion resonances \citep{Moons95, Morbi95a,Morbi97}. Objects may diffuse chaotically from stable to unstable regions \citep{Morbi05}. This process has been shown to occur for Neptune's 3:2 and 2:1 resonances, amongst others \citep{Morbi97,Tiscareno09, Nesvorny002:3, Nesvorny01, deElia08}. Objects leaving mean motion resonances in the Kuiper belt, in this way, may be the main source of Neptune encountering objects at the age of the solar system \citep{duncan95}. Many of these objects are scattered into the inner planetary system, and could be the source of Centaurs or Jupiter Family comets \citep{deSisto10, Morbi97, LevisonDuncan97, Holman93}.

The dynamical processes occurring in the Kuiper belt may well be applicable to exoplanetary systems with a similar structure, \ie an outer planetesimal belt and interior planets. 
The outer belt could be truncated by resonance overlap \citep{Wisdom1980}. Most particles would then inhabit a predominately stable region exterior to this, containing small regions that are unstable due to the overlap of secular or mean motion resonances of the inner planets. Objects could diffuse chaotically on long timescales from the stable to unstable regions and be scattered by the outer planet. Some of these scattered objects could enter the inner planetary system, whilst some could be ejected. 

In our consideration of the dynamics of material scattered from the outer belt by interior planets, we find that this dynamics is strongly dependent on the initial value of the Tisserand parameter, with respect to the outermost planet. Therefore, it is important to consider the value of this parameter. Objects in the outer belt tend to have $T>3$, whilst in order that an object be scattered by the planet, the Tisserand parameter must be less than $3$. Hence, for particles that are scattered at late times due to chaotic diffusion into an unstable region, at the time of first scattering, the Tisserand parameter would be expected to be close to 3. Simulations of our solar system found this to be the case \citep{LevisonDuncan97}. Here, we consider the initial value of the Tisserand parameter of such particles as an unknown, with the expectation that objects scattered in the way described will have initial Tisserand parameter values just below 3.

As a final note, we point out that not all objects scattered by Neptune originate from the cold Kuiper belt. The two other main sources are Neptune's {\it scattered disc} and the Oort cloud. It is possible that similar classes of objects exist in exo-planetary systems, however, there is at present no evidence for exo-Oort clouds or {\it scattered discs}. The distribution of the Tisserand parameter for such objects would differ significantly from those that leave the cold Kuiper belt, in particular for Oort cloud objects, where it is unconstrained and $T<2$ is possible. Therefore, for clarity and simplicity, in this work we focus on the objects that originate in an outer belt and that are first scattered by the outermost planet, at the age of the system. 


\begin{figure*}
\includegraphics[height=50mm]{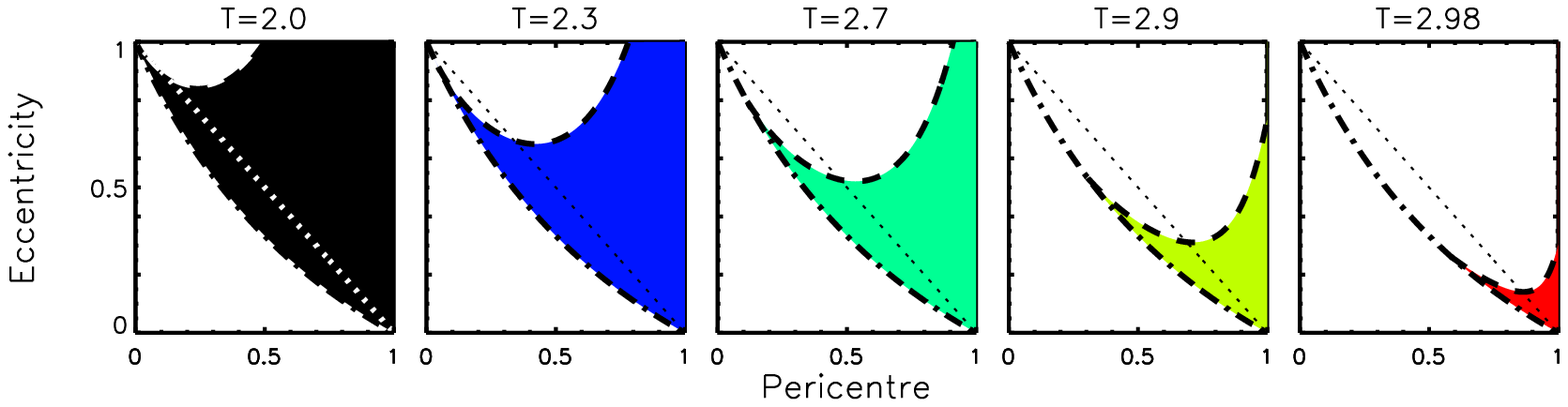}
\includegraphics[height=50mm]{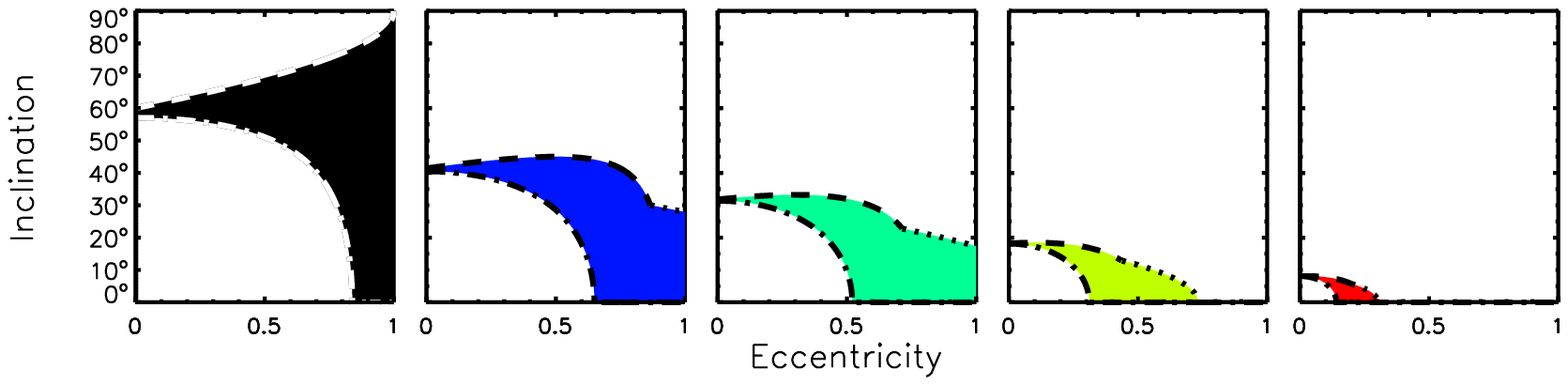}
\includegraphics[height=50mm]{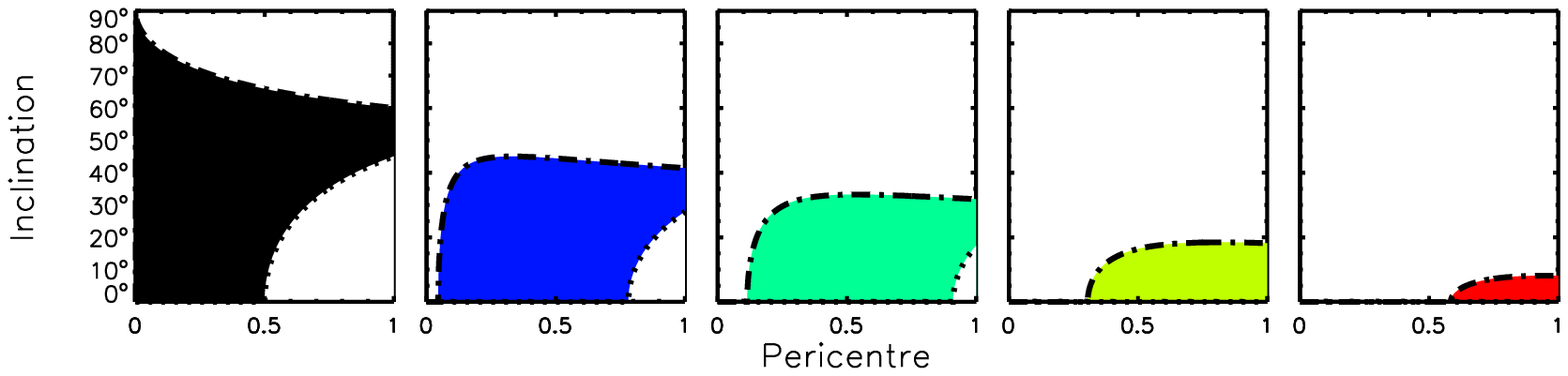}

 \caption{The possible orbital parameters of particles scattered by a single planet, with a given value of the Tisserand parameter with respect to that planet, $T_i$. This forms a 3-D parameter space, that is shown here projected onto the eccentricity-pericentre ($q-e$) plane, the inclination-pericentre ($I-q$) plane and the eccentricity-inclination ($e-I$) plane. The limits of the parameter space are defined analytically in the Table~\ref{tab:bounds}. The units of the pericentre is the planet's semi-major axis and the dotted black line in the top row of plots shows the line where the particle's semi-major axis is equal to the planet's ($a=\frac{q}{1-e}=a_i$). }
\label{fig:param}
\end{figure*}


\section{Scattering by a Single planet}

\label{sec:onepl}
Firstly we consider a system similar to that described in the previous section, with a single planet, labeled by subscript $i$ on a circular orbit at $a_i$ and an exterior planetesimal belt. We consider planetesimals scattered from the outer belt by the planet. We make the simplifying assumption that planetesimals only interact with the planet if their orbits directly cross the planet's orbit. This simplifies the following analysis and enables analytical limits to be easily derived. However, since in reality interactions will occur in a zone around the planet, care should be taken in rigorously applying any of the derived limits, in particular for more massive planets. This will be discussed further in \S\ref{sec:limit}.


\subsection{Orbital constraints}
For a planetesimal with a given value of the Tisserand parameter with respect to this planet, $T_i$, the potential orbits onto which it can be scattered are limited, no matter how many times it interacts with the planet. The Tisserand parameter gives us no information about the probability for any given interaction to scatter a planetesimal onto a given orbit, nor the timescales for interactions to occur. It does, however, limit the orbital parameters of the planetesimals after the interaction, in terms of its pericentre, $q$, eccentricity, $e$ and the inclination, $I$, of its orbit with respect to the planet's. These constraints can be represented by a 3D volume in $(q,e,I)$ space. A planetesimal, given an initial value of $T_i$, may not be scattered onto an orbit with parameters outside of this volume, in this simple example.

This parameter space can be fully mapped out analytically by re-writing Eq.~\ref{eq:tiss} as
\begin{equation}
T_i= \frac{a_i (1-e)}{q} +2\sqrt{\frac{(1+e)q}{a_i}} \cos (I),
\label{eq:t1}
\end{equation}
and noting that if the planetesimal is to remain on a bound orbit, $0<e<1$, $-1<\cos (I)<1$ and $q>0$ must apply. In order that the particle is scattered by the planet its orbit must cross the planet's and thus $Q>a_i$ and $q<a_i$ must apply. Applying these constraints to Eq.~\ref{eq:t1}, places analytical bounds that define this 3D volume of permitted orbits. Given the difficulties in presenting a 3D volume, we instead present the 2D projection of this 3D volume onto the $q-e$ plane, $I-q$ plane and $e-I$ plane, shown in Fig.~\ref{fig:param}. The analytical bounds are presented in Table.~\ref{tab:bounds}.

\subsection{Minimum pericentre}
\label{sec:minq1}
\begin{figure}

\includegraphics[width=0.48\textwidth]{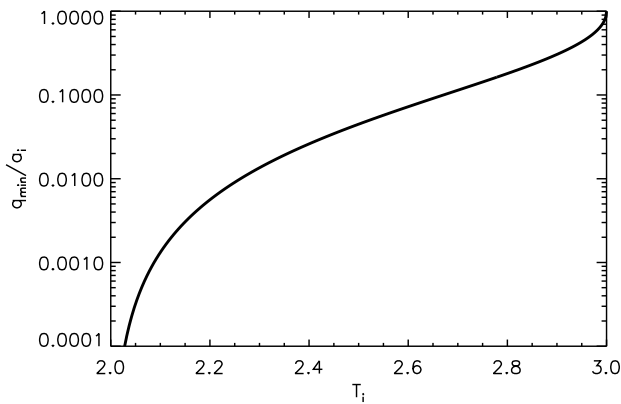}

 \caption{The minimum pericentre for a test particle scattered by a single planet, as a function of the Tisserand parameter value, from Eq.~\ref{eq:minq}. For $T_i<2$, $q_{min}\to 0$.  }

\label{fig:qmin}
\end{figure}

Further examination of the $q$-$e$ plot in Fig.~\ref{fig:param} makes clear that planetesimals cannot be scattered further towards the star than a limiting value, $q_{min}$, determined by T. This value can be calculated using constraints on the orbital parameters, $Q=a_i$ and $\cos (I)=1$ (equivalent to the lower bound in the $q-e$ plane). For $2<T_i<3$:

\begin{equation}
\frac{q_{min} }{a_i}=\frac{-T_i^2 +2T_i +4 - 4 \sqrt {3-T_i}}{T_i^2-8}.
\label{eq:minq}
\end{equation}
$q_{min}$ as a function of $T_i$ is shown in Fig.~\ref{fig:qmin}. The eccentricity at $q_{min}$ will be given by: 

\begin{equation}
e_{lim}=T_i-3 + 2 \sqrt{ 3-T_i}.
\label{eq:elim}
\end{equation}

For $T_i<2$, the lines $Q=a_i$ and $\cos (I) = 1$ (positive root) no longer cross and the parameter space in the q-e plane is no longer bounded by $Q=a_i$, rather by $\cos (I)=1$ (both positive and negative root). Therefore $q_{min} \to 0$. This can be shown to be true by considering the derivatives of the lines:\begin{equation}
\frac{dq}{de}|_{cosI=1, q\to 0} > \frac{dq}{de}|_{Q=a_i, q\to 0} .\nonumber
\end{equation}

Importantly this implies that the constraints on the pericentre that apply to the orbits of objects with $T_i>2$ are not applicable to those with $T_i<2$; such objects can be scattered onto orbits with any pericentre.

\subsection{Ejection}
\label{sec:eject}
A single planet can also eject planetesimals, given a suitable value of the Tisserand parameter. Unbound orbits (\ie those with $e>1$) are not included in the plots in Fig.~\ref{fig:param}. It is, however, possible to determine from the top panel of Fig.~\ref{fig:param} those values of the Tisserand parameter for which the particles are constrained to bound orbits with $e < 1$. The most eccentric orbits are those with pericentre at the planet's orbit ($q=a_i$), therefore substituting into Eq.~\ref{eq:t1}, $(q=a_i, e=1, I=0^\circ)$, we find that there is a limit on the Tisserand parameter such that only objects with $T_i < 2\sqrt2$ can be ejected. This has been previously calculated in, amongst others, \cite{LevisonDuncan97}, using the formulation for the Tisserand parameter of a parabolic orbit. 
 It should, however, be noted that this only applies strictly for low mass planets. As the planet mass is increased, as does the zone of influence of the planet. If we assume that particles can interact with a planet if they are within a distance $\Delta$ from the planet, then a particle on a hyperbolic orbit with $T_p=3$ and $q=\frac{9}{8}a_p$ can still be ejected if $q<a_p+\Delta$. $\Delta$ will be a function of the planet's Hill's radius, $R_H$, for example if $\Delta \sim 2\sqrt{3}R_H $ \citep{Gladman90} Jupiter can eject particles with $T_{Jup}=3$, whilst Neptune cannot.


\begin{table*}
\begin{center}
\begin{tabular}{c c c c c}          
    
   \hline\hline                 
Plane &Line & Constraint& Based on    \\
\hline
$q$-$e$: upper &dashed & $e =1 +2q^3 -qT_i + 2q^{3/2} \sqrt{2+q^3-qT_i}$ &$\cos (I) =\pm 1$ \\
$q$-$e$: lower &dot-dashed & $e=\frac{(1-q)}{(1+q)}$& $Q=1$ \\
$I$-$e$: upper &dashed& $\cos(I) =\frac{ T_i -1-e}{2\sqrt{1-e}}$& $Q=1$ \\
$I$-$e$: upper &dotted &$\cos(I)= \frac{T_i-1+e}{2\sqrt{1+e}}$& $q=1$ \\
$I$-$e$: lower & dot-dashed &$\cos(I) = \frac{T_i^{3/2}}{3\sqrt{3(1-e^2)}}$&$\frac{\partial I}{ \partial q }|_{e,T_i}=0$ \\ 
$I$-$q$: upper & dot-dashed &$\cos (I) = \frac{T_i(1+q)-2}{2\sqrt{2q(1+q)}}$& $Q=1$ \\
$I$-$q$: lower & dotted &$\cos (I) = \frac{T_i}{2\sqrt{2q}}$& $e=1$ \\
\hline\hline   

\end{tabular}

\caption{The analytical boundaries on the parameter space constraining the potential orbital parameters of a particle scattered by a planet, where the initial value of the Tisserand parameter is $T_i$. All units are in terms of the planet's semi-major axis; $a_i=1$. For the cases where more than one limit is stated, the upper of the two applies.}
\label{tab:bounds}
\end{center}
\end{table*}


\begin{figure}

\includegraphics[width=0.48\textwidth]{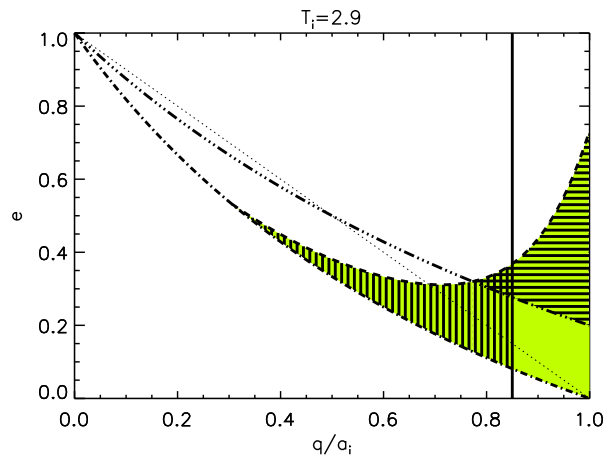}

 \caption{The orbital parameter space as determined by the Tisserand parameter, in the eccentricity-pericentre plane, for $T_i=2.9$ (equivalent to the fourth plot on the top row of Fig.~\ref{fig:param}). The bounds on this space are between the dashed ($\cos I=1$) and dot-dashed lines ($q=a_i$) and shown in green. The subset of this orbital parameter space that can interact with an inner planet placed at $a_{in}=0.8 a_i$ is shown by the vertically hashed region, whilst the subset that could interact with an outer planet placed at $a_{out}=1.5 a_i$ is shown by the horizontally hashed region. The dotted line shows $a=a_i$, the solid line $q=a_{in}$ and the triple dotted dashed line, $Q=a_{out}$.}
\label{fig:param2pl}
\end{figure}


\begin{figure}
\includegraphics[width=0.48\textwidth]{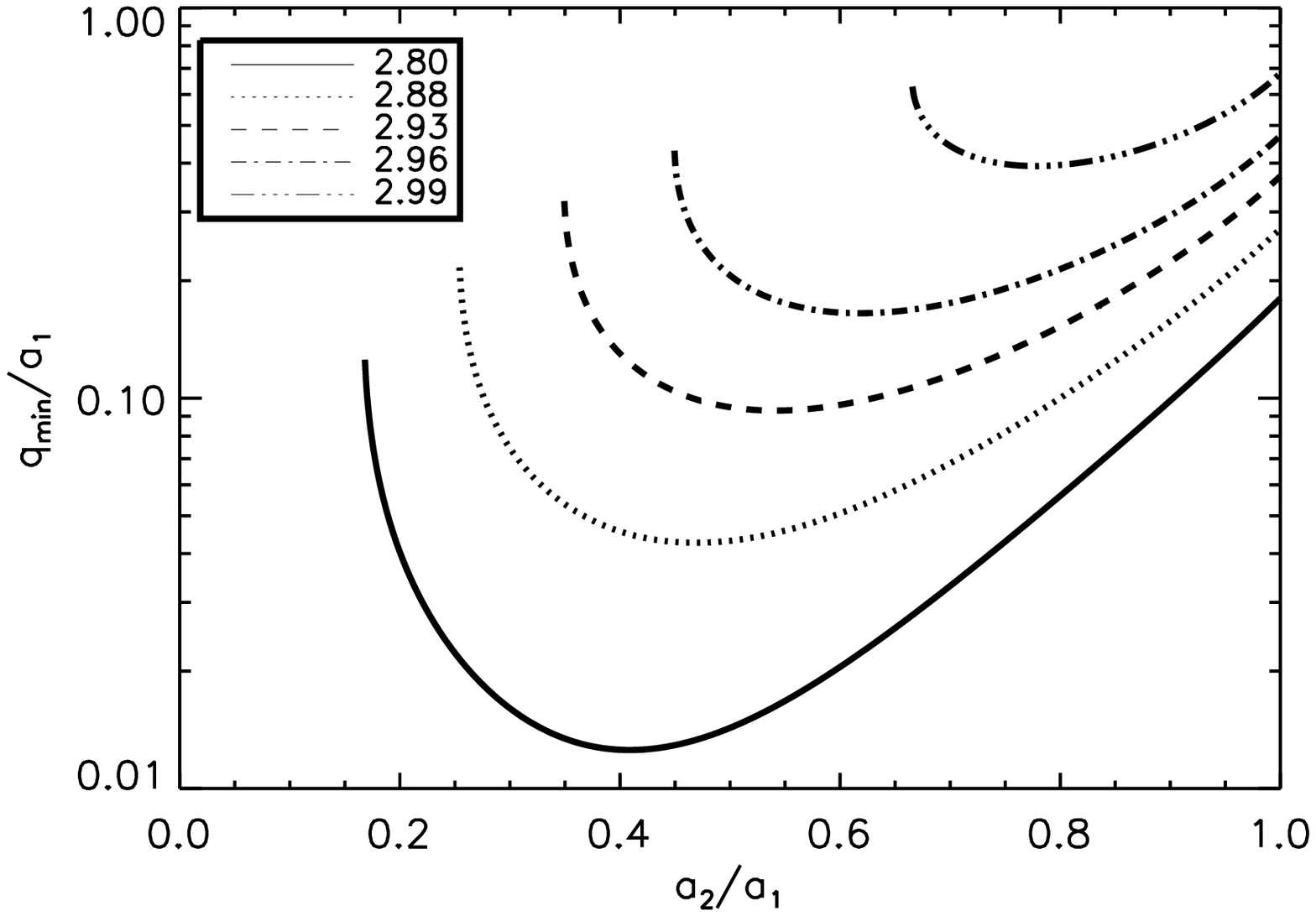}
 \caption{The minimum pericentre for a test particle scattered by two planets, as a function of the ratio of the inner planet's semi-major axis to the outer planet's semi-major axis. }
\label{fig:mina2}
\end{figure}


\section{Scattering by two planets}
\label{sec:twopl}

 Now consider a planetary system with an outer belt and two interior planets, both on circular orbits. 
Particles from the outer belt are scattered by the outer planet, 1. The main possible fates of such particles are ejection, collision with a planet or the star, further scattering interactions with this planet, or scattering by the inner planet, 2. Many scattered particles are scattered multiple times by the outer planet. It dominates their dynamics for a certain period of time, during which the Tisserand parameter, with respect to this planet, $T_1$, is conserved. At some point, the particle may be scattered onto an orbit that overlaps with the inner planet and it may be scattered by that planet. In such an interaction the Tisserand parameter with respect to the inner planet, $T_2$ would be conserved, rather than $T_1$. Depending on the new orbit, it is then likely that the particle is re-scattered by the inner planet and for a certain period its dynamics will be dominated by that planet.

We start by considering this simple situation where the particle is passed from planet 1 to planet 2. This is used to describe constraints on the orbits of scattered particles. We then consider the possibility that particles are scattered backwards and forwards between the two planets in section~\ref{sec:multiscatt}.

\subsection{Orbital constraints} 
\label{sec:orb}

For a particle scattered by the outer planet the Tisserand parameter, $T_1$, is conserved. The value of $T_1$ constrains the orbits, $(q, e, I)$, of scattered particles  to those shown in Fig.~\ref{fig:param} that satisfy Eq.~\ref{eq:t1}. Although only sets of the orbital parameters, $q,e,I$, that satisfy Eq.~\ref{eq:t1} are allowed, the full range of possible values is given by:
\begin{eqnarray}
\nonumber q \in [q_{min} (T_1),  1] \\ \nonumber
 e\in [0, e_{max} (T_1)] \\
I \in [0,I_{max}(T_1)], 
\label{eq:constraints}
 \end{eqnarray}
 where $q_{min}$ is given in Eq.~\ref{eq:minq},  
\begin{equation}
I_{max}= \cos^{-1} (\sqrt{T_1-2}),
\label{eq:imax}
\end{equation}
and 
\begin{equation}
e_{max}=3-T_1 + 2\sqrt{3-T_1}.
\label{eq:emax}
\end{equation}

As mentioned earlier, if $T_1>2\sqrt2$, then $e_{max} >1$ and some orbits are unbound. 

The particle may interact many times with the outer planet, moving between orbits in this parameter set, until at some point it encounters the next planet, 2.
Only a subset of the orbits specified by $T_1$ can interact with the next planet, 2. These are shown by the green filled area in Fig.~\ref{fig:param2pl} and are those orbits that cross the planet's, with $q<a_2$ and: 
\begin{eqnarray}
\nonumber q \in [q_{min} (T_1) ,  a_2] \\\nonumber
e \in [e_{int}  (\frac{a_2}{a_1}),  e_{lim} (T_1) ]\\
I \in [0 , I_{max}],
\label{eq:bound}
\end{eqnarray}

where, 

\begin{equation}
e_{int}= \frac{a_1+a_2}{a_1-a_2}. 
\label{eq:eint}
\end{equation}

If $\frac{a_2}{a_1} <\frac{T_1-2}{4-T_1}$, then the set of orbital parameters with $I=I_{max}$ do not cross the inner planet's orbit. This occurs if the second planet is inside the maximum in $I$ as a function of $q$ that occurs at $q=\frac{T_1-2}{4-T_1}a_1$ (see Fig.~\ref{fig:param}). In which case $I$ is constrained to be less than $I_{int}$ rather than $I_{max}$, where 
\begin{equation}
I_{int}=\cos^{-1} \left(\frac{T_1(1+\frac{a_2}{a_1})-2\frac{a_2}{a_1}}{2\sqrt{2(1+\frac{a_2}{a_1})}} \right).
\end{equation}

 Once the particle is scattered by planet 2, $T_1$ is no longer conserved, instead the value of $T_2$ when the particle is first scattered by planet 2 is conserved. The range of possible $T_2$ values is determined by the initial value of $T_1$ and the planets' orbits, specified by the ratio of the planet's semi-major axes, $\frac{a_2}{a_1}$.

The minimum possible value that $T_2$ can have occurs for particles on orbits with minimum pericentre ($q=q_{min}$), the correspond eccentricity ($e=e_{lim}$) and in the orbital plane of the planets ($I=0^\circ)$. It is given by: 

\begin{equation}
T_{2, min}= \frac{a_2(1-e_{lim})}{q_{min}} + 2\sqrt{\frac{(1+e_{lim})q_{min}}{a_2}},
\label{eq:t2min}
\end{equation}
where $e_{lim}$ (Eq.~\ref{eq:elim}) and $q_{min}$ (Eq.~\ref{eq:minq}) are functions of $T_1$.  

Since the Tisserand parameter ($T_2$) is a monotonically increasing function of $q$, $T_2$ will be maximum for the orbit with the largest value of the pericentre, $q$, that still crosses the planet's orbit, \ie $q=a_2$. For the range of $T_2$ values for orbits with $q=a_2$, the minimum is at $\cos I= \pm 1$ and $e=e_X$, where from the top line of Table~\ref{tab:bounds},
\begin{equation}
e_X= 1 + 2(\frac{a_2}{a_1})^3- (\frac{a_2}{a_1}) T_1 + 2(\frac{a_2}{a_1})^{3/2}\sqrt{2+(\frac{a_2}{a_1})^3-(\frac{a_2}{a_1}) T_1}.
\end{equation}
Hence, the maximum of $T_2$ is given by:
\begin{equation}
T_{2,max}=(1-e_X) + 2\sqrt{1+e_X}.
\label{eq:t2max}
\end{equation}

For the next time period the dynamics of the particle is controlled by the second planet. It may be scattered once or many times. Yet again, the particle's orbit is constrained to orbital parameters, $(q,e,I)$, specified by the value of $T_2$ and Eq.~\ref{eq:t1}. This time, however, we consider the situation where only $T_1$  and the planet's orbits are specified initially such that it is only known that $T_2$ lies between $T_{2, min}$ and $T_{2, max}$. The full range for the orbital parameters $(q, e, I)$ is therefore specified by: 
\begin{eqnarray}
 q_{min} (T_{2,min}) < &q& <  1 \\
 0 <& e &<  e_{max} (T_{2,min}) \\
0 <& I  &< I_{max}(T_{2, min}), 
\label{eq:range}
 \end{eqnarray}
where $q_{min}$ is given by Eq.~\ref{eq:minq}, $e_{max}$ by Eq.~\ref{eq:emax} and $I_{max}$ by Eq.~\ref{eq:imax}, but as a function of $T_{2,min}$ rather than $T_1$.


\subsection{Constraints on which particles interact with the inner planet}

For specific planetary orbits, specified by the ratio of the planets' semi-major axes, $\frac{a_2}{a_1}$, and strict constraints on the initial value of the Tisserand parameter in the outer belt (\ie $T_1$ close to 3), the orbits of scattered particles may be constrained such that they never interact with the inner planet. This occurs when the minimum pericentre to which particles may be scattered by the outer planet is further from the star than the inner planet's orbit; $q_{min} (T_1) >a_2$ (Eq.~\ref{eq:minq}) or :
\begin{equation}
\frac{-T_1^2 +2T_1 +4 - 4 \sqrt {3-T_1}}{T_1^2-8} > \frac{a_2}{a_1}
\end{equation}

\subsection{Minimum pericentre}

In Sec.~\ref{sec:minq1}, Eq.~\ref{eq:minq}, we determined the minimum pericentre to which a single planet may scatter a particle. A similar calculation may be made for two planets, assuming that particles are only passed once along the chain of planets. The minimum pericentre will depend on the Tisserand parameter with respect to the outer planet, $T_1$ and the ratio of the planets' semi-major axes, $\frac{a_2}{a_1}$.

For a particle that is scattered by the outer planet, with a value of the Tisserand parameter with respect to that planet of $T_1$, if it is then scattered by the inner planet, the particle could have a range of possible values of the Tisserand parameter with respect to the inner planet, between $T_{2,min}$ (Eq.~\ref{eq:t2min}) and $T_{2,max}$ (Eq.~\ref{eq:t2max}). Since $q_{min}$ (Eq.~\ref{eq:minq}) is a monotonically increasing function of the Tisserand parameter, the minimum pericentre for scattering by both planets will be given by $q_{min}(T_{2,min})$, where $T_{2, min}$ is the minimum value of the Tisserand parameter (Eq.~\ref{eq:t2min}). If a particle is to eventually be scattered inwards as far as possible by the outer and inner planet, it must be passed from the outer to the inner planet with an orbit of eccentricity $e=e_{lim}(T_{2,min})$ (Eq.~\ref{eq:elim}) and inclination, $I=0^\circ$.

The minimum pericentre for a two planet system is shown in Fig.~\ref{fig:mina2} as a function of the ratio of the planets' semi-major axes, $\frac{a_1}{a_2}$. 
This is calculated from Eq.~\ref{eq:minq}, such that $q= q_{min} (T_{2,min})$, where $T_{2,min}=T (q=a_p, e=e_{lim}(T_1), I=0^\circ)$, using Eq.~\ref{eq:t1}. This has a clear minimum, which occurs at:
\begin{equation}
a_{2,min}= \frac{(1+e_{lim}(T_1))^{1/3} q_{min}(T_1) }{(1-e_{lim}(T_1))^{2/3}},
\label{eq:a2min}
\end{equation}
where $e_{lim}$ and $q_{min}$ are the minimum pericentre and limiting eccentricity for scattering by the outer planet, given by Eq.~\ref{eq:minq} and Eq.~\ref{eq:emax}. 

This means that the optimum configuration of two planets in terms of their ability to scatter particles as close to the star as possible, involves planets positioned in semi-major axis at $a_{2,min}$ and $a_1$. It is interesting to note that the optimum position for the inner planet is not as close to the star as the outer planet could possibly scatter particles \ie $q_{min}(T_1)$, but closer to the outer planet. This is because there is a balance between moving the inner planet closer to the star, decreasing $a_2$, such that $q_{min}$ is decreased directly or moving the planet further from the star, increasing $a_2$, but decreasing $T_2$ and thus $q_{min}$.  Of course this does not include any information about the probability that the particle is ejected or collides with the planet rather than being ejected.

\subsection{Further scattering}
\label{sec:multiscatt}

Scattering is not confined to the forward direction. Particles may originate in the outer belt, be scattered inwards by the outer planet, passed onto the inner planet, and then scattered back outwards again to the outer planet. Constraints on which particles might re-interact with the outer planet can be determined using a similar procedure to that discussed in the previous section (Sec.~\ref{sec:orb}) for particles passed from the outer planet to the inner planet.

The possible values for the orbital parameters of particles scattered by the inner planet are determined by the value of the Tisserand parameter, $T_2$. A subset of these orbits cross the outer planet's orbit, those with apocentres outside of its orbit ($Q>a_1$). For the example of an outer planet at $a_1=1.5a_2$ and with $T_2=2.9$, this subset is shown by the hashed region in Fig.~\ref{fig:param2pl}. Each set of orbital parameters in this region $(q,e,I)$ will specify a possible value for the Tisserand parameter with respect to the outer planet, $T_1$. The minimum possible new value of $T_1$ occurs at the maximum pericentre ($q=a_2$), the maximum eccentricity ($e_{max}(T_2)$ Eq.~\ref{eq:emax}) and $\cos I =1$, such that:
\begin{equation}
T_{1,new,min}= \frac{a_1 (T_2 -2 - 2\sqrt{3-T_2})}{a_2} + 2 \sqrt{\frac{(4-T_2 +2\sqrt{3-T_2})a_2}{a_1}}. 
\label{eq:mint1}
\end{equation}
If there are a range of values for $T_2$, the smallest (\eg $T_{2,min}$ for Eq.~\ref{eq:t2min}) will give the lowest value of $T_{1,new,min}$. The maximum value of $T_1$ such that particles can still interact with the outer planet is 3, as for any scattering event.

 
 If the particle is scattered backwards and forwards multiple times this procedure may be repeated to determine the full range of Tisserand parameter values and potential orbits. $T_{1,new, min}$ can be significantly lower than the initial value of $T_1$ in the outer belt, particularly after multiple scatterings backwards and forwards. Thus, this increases the range of potential orbits of scattered particles.

This can be illustrated using an example system. Consider a particle scattered by the outer planet, with $T_1=2.99$. The inner planet is placed arbitrarily at $a_2=0.7a_1$. The minimum pericentre for the particle after the particle is scattered by both planets, shown in Fig.~\ref{fig:mina2}, is $q_{min}=0.43a_1$. If the particle is then scattered back outwards, the minimum value of $T_1$ is 2.93 (Eq.~\ref{eq:mint1}). If the particle is then scattered back in, again from Fig.~\ref{fig:mina2}, this gives a new minimum pericentre for scattering by the two planets of $q_{min}=0.12a_1$. After a further scattering backwards and forwards, $q_{min} \to 0$; all constraints on the eccentricity and pericentre of the orbit are removed. Given sufficient repetitions this occurs for all pairs of planetary orbits, where the constraints on the Tisserand parameter allow particles to be passed between them. Thus, the orbital parameter space available to scattered particles can be greatly increased by repeatedly scattering them backwards and forwards.

So far we have merely outlined the orbital parameter space available to particles and not discussed the probability for scattering particles into this space. This is in general beyond the scope of this paper, however, these have important implications for the passing of particles backwards and forwards between the two planets. Firstly, it is clear that the timescales for particles to be repeatedly scattered backwards and forwards between two planets will be long and therefore at any given time the probability will be higher that particles have merely been scattered by the outer planet, or passed from the outer to the inner planet once. Secondly, although repeated passing of particles between planets greatly increases the range of orbital parameters available to such scattered particles, this does not mean that it is most probable for such particles to be scattered onto more extreme (higher eccentricity or inclination) orbits. In fact, if we were to assume that a particle has an equal probability of being scattered onto any of the orbital parameters available to it, it is most likely that the particle is scattered onto an orbit that retains a value of the Tisserand parameter close to its original value. It is only the few particles that are scattered onto {\it extreme} orbits, \ie with low pericentre or high eccentricity/inclination, that have significantly reduced values of the Tisserand parameter when they are scattered by the next planet. 
Therefore, although it is possible that particles may be scattered onto extreme orbits, with low values of the Tisserand parameter, by being repeatedly passed backwards and forwards between the planets, we anticipate that the probability for this to occur is low and we are therefore justified in focusing on particles scattered directly along a planetary system for the rest of the paper.

\section{Multi-planet systems}

\label{sec:multi}
All of the calculations discussed so far can be easily applied to planetary systems with many planets. The procedure discussed in Sec.~\ref{sec:orb} can be repeated many times, to determine the full range of orbital constraints and values for the Tisserand parameter after scattering by each planet. 
This analysis places useful constraints on the planets with which particles can interact, the planets that can eject particles and the minimum pericentre to which the whole system can scatter particles.

All of the dynamics is determined by the initial value of the Tisserand parameter with respect to the outer planet, $T_1$, the outer planet's semi-major axis, $a_1$ and the ratio of the planets' semi-major axes to one another, $\frac{a_{i+1}}{a_i}$. Scaling the system, \ie changing the semi-major axes, $a_i$, whilst keeping their ratios, $\frac{a_{i+1}}{a_i}$, constant, will not affect the dynamics (values of $T_i$) and merely scales the minimum pericentre, $q_{min}$, with $a_1$. In the next section, we discuss these constraints in terms of an example planetary system.

\begin{figure}
\includegraphics[width=0.48\textwidth]{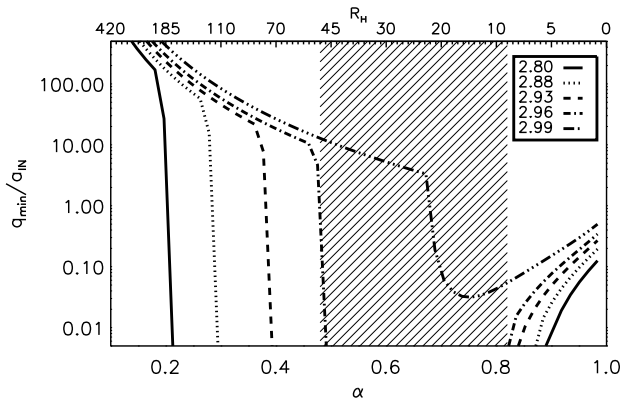}
\caption{The variation in the minimum pericentre to which test particles can be scattered to by a system of five planets. The ratio of the planets' semi-major axes ($\alpha$) is constant and is given as a ratio on the bottom axis and in terms of separation in Hill's radii, for five 10M$_{\oplus}$ planets, on the top axis. The initial value of the Tisserand parameter with respect to the outer planet is varied between 2.8 and 3.0.  The shaded region illustrates the ``unconstrained'' regime for particles with $T=2.96$, whilst the region to its left is the ``non-interacting''  and the region to its right is the ``constrained'' regime (see discussion in text).}
\label{fig:chain}
\end{figure}

\subsection{A hypothetical 5-planet system with constant ratio of planets' semi-major axes}
~\label{sec:alpha}
We apply these calculations to a system of 5 planets, separated by a constant ratio of adjacent planets' semi-major axis ($\frac{a_{i+1}}{a_i}= \alpha$). This corresponds to a constant number of Hill's radii for equal mass planets. Our results are independent of the planet masses. We fix the inner planet at $a_5=a_{IN}$ and calculate the semi-major axes of the other planets accordingly for a range of values for $\alpha$.

The minimum pericentre to which this system can scatter particles, shown in Fig.~\ref{fig:chain} as a function of $\alpha$, is calculated by repeatedly determining the minimum value of the Tisserand parameter for each planet. For the $i$th planet this occurs at $q=q_{min}(T_{i+1, min})$ (Eq.~\ref{eq:minq}), $e=e_{lim}(T_{i+1,min})$ (Eq.~\ref{eq:elim}) and $\cos I =1$.  

In this plot scattered particles exhibit three types of behaviour. For simplicity we label the three types of behaviour as ``non-interacting'', ``constrained'' and ``unconstrained''. This refers to the constraints on the orbits of scattered particles. In the ``non-interacting'' regime, the planets are so widely separated (small $\alpha$) that particles cannot be scattered all the way along the chain of planets. The minimum pericentre to which one of the planets can scatter particles is outside of the next innermost planet's orbit. Hence the particles are restricted to the region surrounding the outer planet(s). 

In the ``constrained'' regime, the planets are so close together (large $\alpha$) that particles can be scattered between all planets in the system. If they are only scattered once along the chain of planets, the Tisserand parameter cannot vary significantly from its original value and there will be a non-zero minimum pericentre to which particles can be scattered. For such closely separated planets, it may no longer be valid to treat the scattering as a series of three body problems and the probability that particles are passed backwards and forwards between planets increases. This and the stability of planets so close together question whether particles scattered in any planetary system actually exhibit behaviour reminiscent of this ``constrained'' regime.

As the separation of the planets is increased, the minimum possible value of the Tisserand parameter for each planet decreases and hence the minimum pericentre for the whole system decreases. Eventually the separation is large enough that the Tisserand parameter falls below 2 and all constraints on the minimum pericentre are removed. This forms the third, ``unconstrained'' regime, where there are few constraints on the orbital parameters of scattered particles.

\begin{figure}
\includegraphics[width=0.48\textwidth]{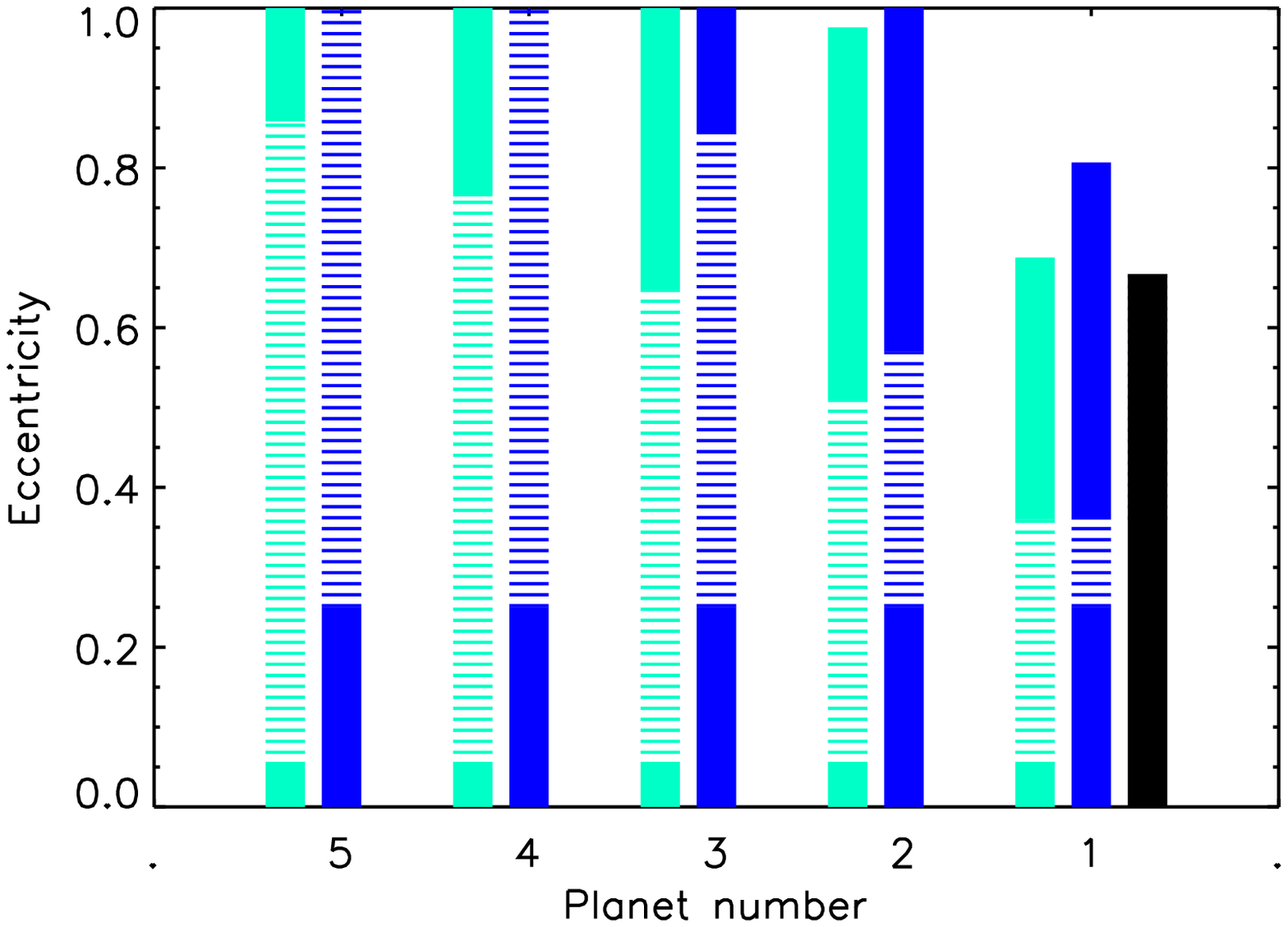}
\includegraphics[width=0.48\textwidth]{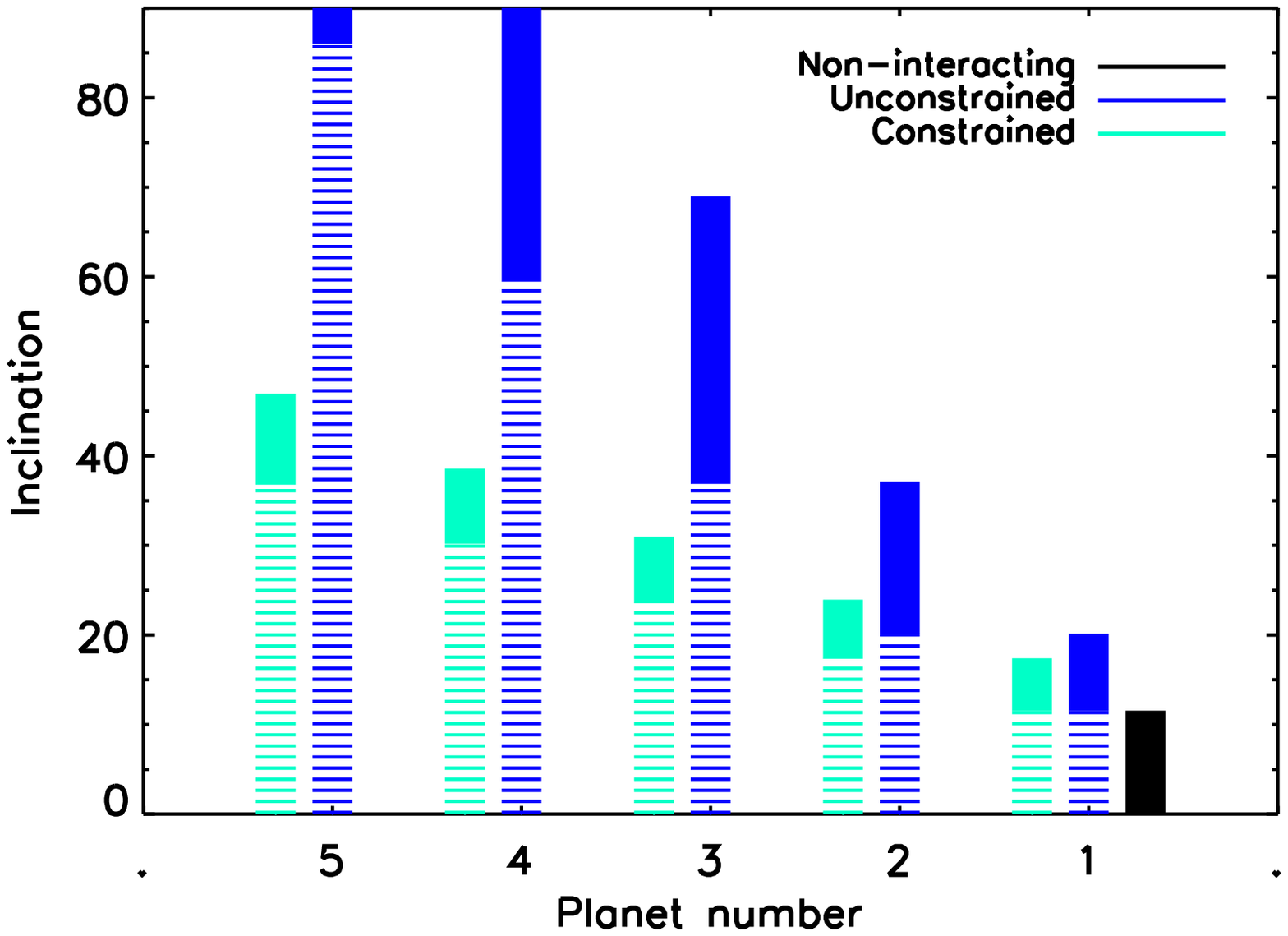}

\caption{Constraints on the eccentricities and inclinations (Eq.~\ref{eq:range}) of particles scattered by a system of five planets with constant ratio of the planets' semi-major axes, $\alpha$, and an initial value of the Tisserand parameter in the outer belt of $T_1=2.96$. Particles are scattered from the belt, outside of planet 1, to the innermost planet, 5. Three planet separations are considered, corresponding to the three regimes (see discussion in the text); ``non-interacting'', $\alpha=0.2$, ``unconstrained'', $\alpha=0.6$ and ``constrained'', $\alpha=0.9$. The dashed regions correspond to the parameters of particles that can interact with the next interior planet (Eq.~\ref{eq:bound}).  The particles with high eccentricity were scattered outwards and therefore are not on orbits that cross the inner planet's orbit. }
\label{fig:alpharange}
\end{figure}

In Fig.~\ref{fig:alpharange} the constraints on the eccentricities and inclinations of particles in the 3 regimes are shown. As particles are scattered by each planet, from the outermost (1) to the innermost (5), there will be a range of possible Tisserand parameter values, between $T_{i,min}$ (Eq.~\ref{eq:t2min}) and $T_{i,max}$ (Eq.~\ref{eq:t2max}) and hence a range of possible orbital parameters, given by Eqs.~\ref{eq:range}, although of course the approximations used to calculate these may mean that they are not always strictly applicable, see discussion in \S\ref{sec:limit}. It is the maximum inclination and eccentricity that are important on this figure, although of course the orbits of scattered particles will be distributed between the minimum and maximum values, in a manner not determined by this analysis. The plot shows that, for this example with $T_1=2.96$, almost all planets can eject particles ($e>1$) and that the scale height of the disc (inclinations of scattered particles) increases with decreasing distance to the star, as the constraints on the orbits of scattered particles decrease with each successive scattering event. It is clearly seen, as anticipated, that the constraints of orbits in the ``constrained'' regime are much tighter than those in the ``unconstrained'' regime.

Although very few real planetary systems have planets separated by a constant ratio of their semi-major axes, it may be possible to similarly classify the behaviour of scattered particles within the three regimes and thus usefully better understand the future fate of scattered particles.

\begin{figure}
\includegraphics[width=0.48\textwidth]{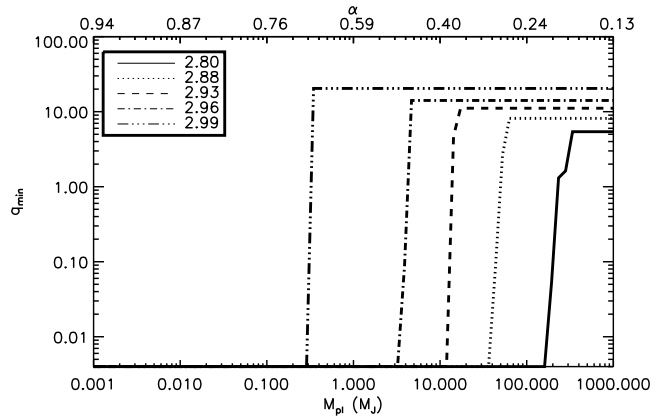}
\caption{The same as Fig.~\ref{fig:chain}, but for tightly packed planetary systems, with equal mass planets separated by $10R_H$. The mass of the planets is shown on the bottom axis, whilst the top axis shows $\alpha$. As many planets as fit between 1 and 30AU are included, hence the minimum pericentre is no longer finite for large $\alpha$.   }
\label{fig:alpha:manypl}
\end{figure}

\subsection{Hypothetical multi-planet system separated by $10R_H$}
For real planetary systems the planets cannot be arbitrarily close together as dynamical instabilities are important. \cite{Chambers96} find that planets must be separated by at least $10R_H$ to be stable. On Fig.~\ref{fig:chain}, the separation of the planets is shown in terms of Hill's radii on the top axis, for a system of equal mass $10M_{\oplus}$ planets. This shows that for the 10$M_{\oplus}$ planets considered, if they are separated by $10R_H$, then the behaviour of particles is unconstrained ($q_{min} \to 0$). Only very low mass ($<10M_{\oplus}$) systems may be dynamically stable (separated by more than $10R_H$) and have limits on the scattering of particles, such that the particle's behaviour is in the ``constrained'' regime. 

Such low mass systems are, however, unlikely to only contain 5 planets. One possible outcome of planet formation, is a chain of low mass planet embryos and an outer disc of planetesimals. Consider the example of such a disc in the position of the Solar System's Kuiper belt and a chain of interior, equal mass planets, between 1 and 30AU. If planets generally form on orbits as tightly packed as possible \citep{Barnes04, Raymond09}, then their separation will be $\sim 10 R_H$. We investigate the dynamics in such a system by varying the planet mass and thus the number of planets that fit between 1 and 30AU. This is equivalent to varying $\alpha$. The results are shown in Fig.~\ref{fig:alpha:manypl}. The behaviour is identical to the five planet system in the ``non-interacting'' and ``unconstrained'' regimes, however the ``constrained'' regime no longer exists. This is because the interior planets further increase the parameter space available to scattered particles. 


\section{Applications to real systems}
\label{sec:application}
\subsection{Solar System}

\label{sec:ss}

This analysis can be applied to the planetary system that we understand best, our Solar System. There are three possible sources of scattered bodies; the asteroid belt, the Kuiper belt and the Oort cloud.
Most of the discussion so far has applied to the scattering of particles from a Kuiper-like belt, however, very similar processes occur in the asteroid belt. As discussed in \S\ref{sec:initial}, we anticipate that the Tisserand parameter for objects scattered out of the Kuiper belt is close to 3. This should also apply to asteroids scattered from the main belt by Mars. The main difference between scattered asteroids and scattered Kuiper belt objects will be in the distribution of the Tisserand parameter, $T_{Mars}$ or $T_{Nep}$. This work does not determine these distributions, however, we speculate that the distribution of $T_{Mars}$ may be spread to lower values than that of $T_{Nep}$. Jupiter is a strong perturber and may be able to alter the orbital parameters of an asteroid significantly in a single encounter. 
For Oort cloud comets scattered by planets, on the other hand, the range of values of the Tisserand parameter is even larger, potentially including many comets with $T_p<2$. This means that although this analysis is most usefully applied to scattered Kuiper-belt objects, it can also be applied to scattered asteroids, but it cannot place many limitations on the orbits of scattered Oort cloud comets.

\begin{figure}
\includegraphics[width=0.48\textwidth]{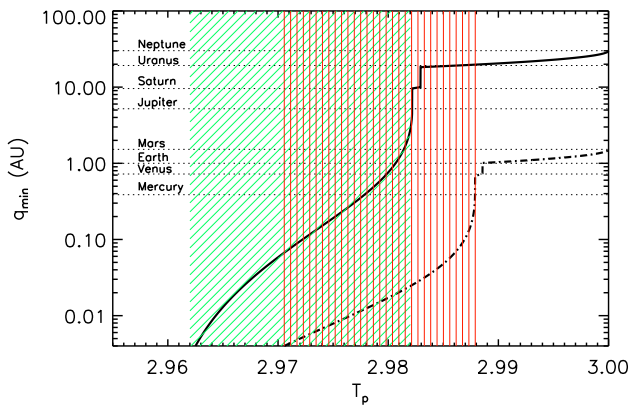}

\caption{ The minimum pericentre ($q_{min}$) to which Kuiper belt objects can be scattered to by Neptune and the outer Solar System planets (solid line) and similarly for asteroids scattered by the terrestrial planets (dashed line), as a function of the initial value of the Tisserand parameter with respect to Neptune or Mars, respectively. The shaded area illustrates the ``constrained'' region, for scattered Kuiper belt objects with diagonal shading in green and for scattered asteroids with vertical shading in red. The ``unconstrained'' region lies to the left of the plot and the ``non-interacting'' regime to the right.   }
\label{fig:qmint}
\end{figure}

 Firstly considering objects scattered out of the Kuiper belt. Useful constraints can be made on their orbits in the outer planet region using this analysis; \eg particle inclinations are constrained to be below a maximum value, for example $80^\circ$, for scattered Kuiper belt objects with $T_{Nep}>2.96$, consistent with observations of Centaurs \citep{inclination}. It can also be inferred that the Solar System's outer planets are well placed for scattering particles between them. If $T_{Nep}\le 2.982$ then particles can be scattered, directly, all the way along the chain of planets to Jupiter and Table~\ref{tab:ss} shows that using Eq.~\ref{eq:a2min} the planets are placed close to optimally for scattering particles as far inwards as possible. The three regimes presented in \S\ref{sec:alpha} can be applied to the Solar System to show that the majority of scattered Kuiper belt objects exhibit behaviour consistent with the ``unconstrained'' regime. These are shown in Fig.~\ref{fig:qmint}. If $T_{Nep}<2.962$, there are no constraints on the orbits of scattered particles inside of Jupiter. On the other hand, if $2.962 < T_{Nep} < 2.982$ the dynamics of scattered bodies are ``constrained'' and there is a minimum value for the pericentre of objects scattered by Jupiter, whilst if $T_{Nep}>2.982$ the particles are ``non-interacting'' and cannot be scattered into the inner planetary system.

 Similar constraints can be made for asteroids scattered into the inner planetary region. If $T_{Mars}<2.97$ then scattered objects behave as if they were in the ``unconstrained'' regime and if $T_{Mars}=2.98$, Eq.~\ref{eq:a2min} can be used in a similar manner to determine that the terrestrial planets are close to optimally separated for scattering particles between them, see Table~\ref{tab:ss}. The important question, therefore, is what fraction of scattered of asteroids have $T_{Mars}\sim 2.98$. If $T_{Mars}$ is lower then issues arise with some of the approximations used in this work. For example, particles may not be scattered directly along the chain of planets. For $T_{Mars}<2.96$, the minimum pericentre to which asteroids may be scattered by Mars is already inside of Venus's orbit, $q_{min}<a_{Venus}$ (Eq.~\ref{eq:minq}). Particles may interact directly with Venus before being scattered by Earth, or be scattered multiple times by both Venus and Earth. As described in \S\ref{sec:multiscatt} this would greatly increase the range of potential orbits for scattered particles. Also mean motion resonances and secular effects will play a greater role in altering the dynamics of scattered asteroids than of scattered Kuiper belt objects, raising further doubt as to the validity of this approach. It can nonetheless be used cautiously to further investigate the dynamics of asteroids scattered into this region.


\begin{table}
\begin{center}
\begin{tabular}{c c c }          
       \hline\hline                 
Planet & \multicolumn{2}{|c|}{Semi-major axis (AU) } \\
& Observed & Optimum \\
 
\hline
\hline

Neptune &30.1 & \\
Uranus & 19.2 & 21.1 \\
Saturn & 9.58 & 10.5 \\
Jupiter & 5.20 & 3.3\\

\hline\hline

Mars & 1.52 & \\
Earth & 1.00& 1.18\\
Venus &0.72 &0.74 \\
Mercury & 0.39 &0.33\\

\hline\hline  
 
\end{tabular}

\caption{ The semi-major axes of solar system planets, compared the optimum semi-major axes in terms of scattering particles inwards. These were calculated using (Eq.~\ref{eq:a2min}) for objects scattered from the Kuiper belt by Neptune with $T_{Nep}=2.98$ and separately for objects scattered from the asteroid belt by Mars, with $T_{Mars}=2.98$. } 
\label{tab:ss}
\end{center}
\end{table}


\subsection{Warm dust discs}


As discussed in the Introduction, there are many observations of stars with {\it warm} dust belts, \eg \citep{Gaidos99, Beichman05, Song05}. Many of the systems with warm dust also have cold dust belts, amongst others, $\eta$ Corvi \citep{resolveHD69830, Wyattetacorvi}, $\beta$Leo \citep{Churcher11} and $\epsilon$ Eri \citep{Backman09}. The analysis presented here can be used to consider the scattering of particles from an outer belt inwards, as a potential explanation for the observed {\it warm} belts. Our main conclusion is that the architecture of a planetary system determines whether or not material can be scattered to the position of the observed belt. We, therefore, speculate that the diversity of planetary system architectures could result in the diversity of observed systems, both in terms of disc radii and the ratio of the flux from the outer to the inner belt. Although this analysis does not determine what fraction of the scattered particles end up in the position of the observed disc, it does show that some planetary systems cannot scatter particles onto the required orbits and illustrates that when the distribution of scattered particles is determined, tight constraints will be placed on the architecture of the planetary system required.

Consider the example system of $\eta$ Corvi, with cold and warm dust. The inner belt is resolved and lies between 0.16-2.98 AU \citep{resolveHD69830}, whilst the outer dust is at 150 $\pm$ 20 AU \citep{Wyattetacorvi}. Although there are no planets detected in this system, it seems probable that there is a planet close to the inner edge of the cold, outer belt, that truncates it \citep{Wyattetacorvi}. We, therefore, consider a planet at 100AU. If the Tisserand parameter with respect to this planet is $T_1=2.96$, then this planet alone could potentially scatter particles in as far as 47AU (Eq.~\ref{eq:minq}). In order for particles to be scattered inwards to the location of the warm belt, $q_{min}<3$AU, at least three planets are required. The optimum position for these planets is 58 and 23 AU, with the outer planet at 100AU (Eq.~\ref{eq:a2min}). The orbits cannot vary significantly from these values if the minimum pericentre is to remain less than 3 AU. For example, if the planets were positioned at 100, 80 and 60 AU, particles could only be scattered in as far as 6AU and thus the warm dust belt, if it formed, would be at larger radii. Alternatively, there could be more than 3 planets, the initial value of the Tisserand parameter could be less than 2.96 or particles could be scattered multiple times backwards and forwards between the planets, as discussed in \S\ref{sec:multiscatt}.

\subsection{Metal polluted white dwarfs and white dwarfs with close-in circumstellar discs }

Evidence of evolved planetary systems and scattering of planetary material is found in the observations of metal polluted white dwarfs \citep{Zuckerman03, Koester05} and white dwarfs with close-in circumstellar discs \citep{farihi09}. In order to explain these observations with planetary material, comets or asteroids must be scattered onto star-grazing orbits and tidally disrupted. The analysis presented in this work can be used to determine the feasibility of this explanation.

Planets are required to scatter comets or asteroids close enough to the star. There are three potential reservoirs in an evolved planetary system, a Kuiper belt analogue, an Oort cloud analogue and if it survives an asteroid belt analogue. This analysis shows that it is possible for particles from all three reservoirs to be scattered onto star-grazing orbits, but that this ability depends strongly on the planets' orbits and the initial value of the Tisserand parameter. The lower the initial value of the Tisserand parameter, the more likely that particles can be scattered sufficiently close to the star (the lower $q_{min}$ Fig.~\ref{fig:qmin}). Hence, the majority of particles from an Oort cloud analogue can be scattered onto star-grazing orbits, whilst for a Kuiper or asteroid belt analogue this ability is strongly dependent on the initial value of the Tisserand parameter and the planets' orbits. Asteroid belt analogues have the advantage of lower initial values for the Tisserand parameter, but the disadvantage that there may be fewer surviving interior planets and the asteroid belt itself may not survive until the white dwarf phase.

There are a large number of observations of Kuiper belt analogues around main sequence stars \citep{wyattreview} and models find that such systems survive the star's evolution \citep{Bonsor10}. Such belts have been suggested as the source of the metal pollution \eg \cite{Jurasmallasteroid}, although there is little evidence that they are capable of scattering particles sufficiently close to the star. Here, we show that it is possible for some planetary systems to scatter particles from an outer belt onto star-grazing orbits, but that there are tight constraints on the planets' orbits and the initial value of the Tisserand parameter in the outer belt.

One potential hindrance in the ability of an evolved planetary system to scatter particles onto star-grazing orbits is the absence of inner planets due to the star's evolution. \cite{villaverlivio} find that white dwarfs should not possess planets within 15AU due to a combination of the increased stellar radius, tidal forces and stellar mass loss. In order for a planet at $a_i=15$AU to scatter particles onto star-grazing orbits ($q_{min}<R_{\odot}$), particles must have values of the Tisserand parameter less than 2.05 when they interact with the planet (Eq.~\ref{eq:minq}). Only particles from an evolved Oort cloud might have sufficiently low values of Tisserand parameter without interacting with further planets. Therefore, using repetition of the technique described in \S\ref{sec:orb}, if particles originate in an outer belt with $T_1>2.97$, then at least 4 planets are required to scatter particles onto star-grazing orbits, whilst for $2.89<T_1<2.97$ only 3 are required.
Another potential hindrance is the instability of many planetary systems after stellar mass loss on the giant branches \citep{DebesSigurdsson}, if, for example, planets are ejected.
Examples of real planetary systems that could scatter particles onto star-grazing orbits from a Kuiper-like belt include our Solar System (if $T_{Nep}<2.96$) and HR 8799 with planets at 14.5, 24, 38 and 68 AU \citep{hr8799detection08, 4thplanethr8799}, if $T_1<2.95$ in the outer belt.

This analysis crucially shows that it is possible to scatter comets or asteroids onto star-grazing orbits and places limits on the architecture of a planetary system that can do this, but it does not inform us about the probability of a given planetary system to scatter planetesimals onto star-grazing orbits. Oort cloud analogues only require a single planet to scatter material onto star-grazing orbits, whilst constraints are placed on the orbits of planets and the initial value of the Tisserand parameter required to scatter material inwards from a Kuiper or asteroid belt analogues. Thus, this analysis shows that material from an evolved Kuiper belt is a potential origin of the metal pollution in white dwarfs, although fewer constraints exist on the ability of an evolved Oort cloud to scatter comets onto star-grazing orbits. This provides important evidence in support of the planetary origin for the white dwarf observations.

\section{Discussion of limitations}
\label{sec:limit}

The purpose of this work is to present a simple analytical tool that can be applied to many planetary systems. It determines the potential orbital parameters of scattered particles, based on the initial value of the Tisserand parameter and the planets' orbits. It does not claim to determine the probability for any particle to be scattered onto a given orbit, nor the expected distribution of scattered particles. In order to retain this simplicity it was necessary to make several assumptions that strictly limit the applicability of this analysis. These are discussed below.

We anticipate that the behaviour described in this work will be useful in interpreting N-body simulations of small body scattering in planetary systems. In such simulations the initial value of the Tisserand parameter for individual particles will be well constrained, and this means that it should be possible to describe their subsequent evolution using the analysis presented here. However, this presupposes that our simplifying assumptions are valid, and it will be important to use N-body simulations to test this. The main assumptions regarding the dynamics that we see are summarised here.

One of the biggest limitations in the analysis presented here, is the dependence on the initial value of the Tisserand parameter in the outer belt. This is in general an unknown quantity, although good approximations can be made to its value, as discussed in \S\ref{sec:initial} and its value is determined in N-body simulations.

Here, the scattering of particles by a chain of planets is considered as a series of three-body problems. This should be broadly true, although some particles may be affected by secular or resonant perturbations, or interact with a planet other than the one dominating their dynamics during that period. This could alter the value of the Tisserand parameter. Particles may also be passed backwards and forwards along the chain of planets, as discussed in \S\ref{sec:multiscatt}.

\begin{figure}
\includegraphics[width=0.48\textwidth]{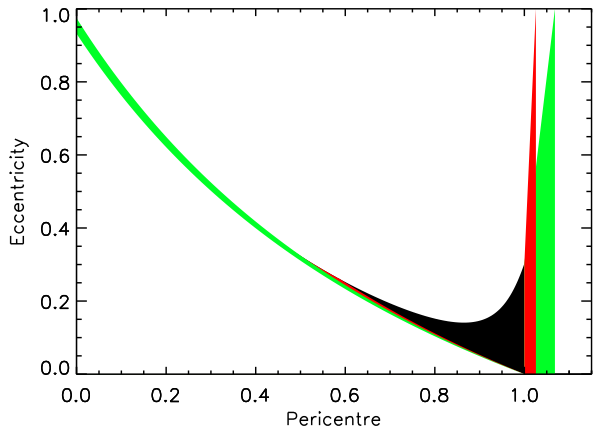}
\caption{ The extension of the orbital parameter space available to scattered particles with $T_p=2.98$. This is analogous to the plots in the top row of Fig.~\ref{fig:param}, except that the particles are allowed to interact with the planet within a zone of size $\Delta \sim R_H$, for a Jupiter mass (green) and Neptune mass (red) planet. For the Neptune mass planet the orbital parameter space available to scattered particles is not altered significantly from that shown in Fig.~\ref{fig:param}, whereas for the Jupiter mass planet the strict limit on the minimum pericentre is removed.  }
\label{fig:extend}
\end{figure}

Another very important limiting assumption is that particles only interact with the planet when their orbits exactly cross the planet's orbit, whereas in reality there will be a small zone around the planet within which a particle may be scattered by the planet.  The size of such a zone depends on the planet's mass and therefore its inclusion into the calculations would introduce a mass dependence to the analysis presented here. For the low mass planets all of the analysis presented here should be valid, whilst for higher mass planets care must be taken when rigorously applying some of the derived limits. This is illustrated in Fig.~\ref{fig:extend}, which shows the extension to the orbital parameter space for particles that can interact with the planet in a zone of size $\frac{\Delta}{a_p}$. This extends the range of potential orbital parameters of scattered particles, as orbits with $Q>a_p(1-\Delta)$ and $q<a_p(1+\Delta)$ can interact with the planet, rather than just $Q>a_p$ and $q<a_p$. This orbital parameter space is shown in Fig.~\ref{fig:extend} for Jupiter and Neptune, with $\Delta =R_H$ and $T_p=2.98$. Jupiter is massive enough that there is no longer a limit on the minimum pericentre to which particles can be scattered and the available parameter space is increased by a small, but significant, amount. Neptune, on the other hand, does not significantly extend the parameter space available to scattered particles and therefore all of the analysis presented in this paper will apply. For Jupiter mass planets the broad results presented here can still be applied, however, care should be taken when rigorously applying the derived limits.



Strictly the conservation of the Tisserand parameter, and therefore this analysis, should only be applied to systems with co-planar planets on circular orbits, \ie within the context of the circular restricted three body problem. It is, however, found that even when these assumptions are relaxed, the analysis still applies approximately, for example \cite{Murray&Dermott} found only a small change in the Tisserand parameter when they consider Jupiter's eccentricity. Caution should, however, be exerted when applying this analysis to some of the detected exoplanets with large eccentricities (and relative inclinations). 



\section{Conclusions}

The purpose of this work was to describes simply and analytically the scattering of particles in any planetary system. This analysis constrains the outcomes of scattering events, based on the conservation of the Tisserand parameter (Eq.~\ref{eq:tiss}), in a manner that is very useful for analysing the structure of many planetary systems where the scattering of small bodies by planets is important. This analysis can be used to better understand behaviour seen in N-body simulations.

We consider here the application to planetary systems where small bodies are scattered from an outer belt by interior planets. The analysis could, however, easily be reformulated to consider scattering by planets exterior to the belt. We assume that the scattering process can be approximated by a series of three-body problems, during each of which the Tisserand parameter with respect to the relevant planet is conserved. This constrains the possible orbits of scattered particles, based solely on the initial value of the Tisserand parameter and the orbits of the planets, with the assumption that particles are passed directly along the chain of planets  and that particles only interact with the planet when their orbits directly cross the planet's orbit. In this case there is no dependence on the planet's mass and it is only the ratio of the planets' semi-major axes that are important. A dependence on planet mass would be introduced if the interactions of particles with the planet within a zone around the planet were included. This analysis places an important limit on how far in particles can be scattered ($q_{min}$ from Eq.~\ref{eq:minq}) and determines which planets the particles can interact with, which can eject them and the potential height of the disc, based on the maximum particle inclinations (Eq.~\ref{eq:range}). We consider the full range of possible orbits of scattered particles, rather than their distribution. 

In this work we consider the application of this analysis to our Solar System, main sequence stars with both cold and warm dust belts and metal polluted white dwarfs. In the Solar System, this analysis describes the scattering of Kuiper belt objects by Neptune to become Centaurs and Jupiter Family comets, as well as asteroids by Mars and the terrestial planets. We show that the Solar System planets are close to optimally separated for scattering particles between them. One explanation for main sequence stars with {\it warm} dust belts that cannot have survived for the age of the system in their current positions is that material was scattered inwards from an outer belt. If this is the case, this analysis shows that certain architectures for the planetary system could not produce the observations. Given the strong dependence of the scattering process on planetary system architecture, we speculate that the diversity of such systems is a reflection of the variety of planetary system architectures.

Observations of metal polluted white dwarfs and white dwarfs with circumstellar discs have been associated with material scattered inwards from an outer evolved planetary system. Such material can be scattered sufficiently close to the star from all Oort cloud analogues and, given certain constraints on the system architecture, from some Kuiper belt analogues. This strengthens the case for a planetary origin to these observations, although this analysis does not comment on the probability for particles to be scattered onto such orbits. 

In summary, the analytical tool presented here can aid our understanding and place useful constraints on the scattering of small bodies in a wide range of planetary systems.

\section{Acknowledgements}
We thank the referee, R. Brasser, for very useful comments that contributed to this paper. We thank D. Veras, G. Kennedy and A. Mustill for comments that contributed to the forming of this manuscript. AB thanks STFC for a student grant.

\bibliographystyle{mn.bst}

\bibliography{ref}

\end{document}